\documentclass[11pt,a4paper]{article}
\usepackage{bm, amssymb,pifont,cancel,amsmath,comment,authblk}
\usepackage[dvips]{graphicx}
\usepackage[colorlinks=true,link color=blue,cite color=blue,urlcolor=black]{hyperref}
\usepackage{braket}

\newcommand{\beq}{\begin{equation}}
\newcommand{\eeq}{\end{equation}}

 \makeatletter
    
    \@addtoreset{equation}{section}
  \makeatother

\begin{document}

\begin{center}
{\Large{\bf{Note on ``Equivalence Between Different Auxiliary Field Formulations\\
 of ${\cal N}=1$ Supergravity Coupled to Matter"}}}
\vskip 6pt
\large{Shuntaro Aoki}$^{1}${\renewcommand{\thefootnote}{\fnsymbol{footnote}}\footnote[1]{E-mail address: shun-soccer@akane.waseda.jp}} and \large{Yusuke Yamada}$^{2}${\renewcommand{\thefootnote}{\fnsymbol{footnote}}\footnote[2]{E-mail address: yusukeyy@stanford.edu}}\\
\vskip 4pt
$^1${\small{\it Department of Physics, Waseda university,}}\\
{\small{\it Tokyo 169-8555, Japan}}
\vskip 1.0em

$^2${\small{\it Stanford Institute for Theoretical Physics and Department of Physics,\\
Stanford University, Stanford, CA 94305, U.S.A.}}\\
\vskip 1.0em
 
\end{center}
In Ref.~\cite{V1}, we claimed that the off-shell supergravity(SUGRA) formulations with different auxiliary fields can be unified with U(1) gauge symmetry, whose gauge superfield does not have a kinetic term. However, after the submission, we found a critical error in our statement.  

As shown in Sec.~3.2 of Ref.~\cite{V1}, we can formally introduce gauged R-symmetry (without gauge kinetic function) into the a general old-minimal SUGRA action as Eq.~(40). This action can be dual to the new-minimal SUGRA, and indeed we can formally introduce a linear superfield by the procedure shown in Ref.~\cite{Ferrara:1983dh}. However, the dual action of that in Eq.~(40) generically has $V_R$ only in a term $[g_RL_0V_R]_D$, where $L_0$ is a real linear compensator, which is shown in Appendix. Since $V_R$ is linear in this term, its equation of motion leads to $L_0=0$, which shows that we cannot identify $L_0$ as a compensator field. For example, in Eq.~(51), we actually find that $V_R$ has only a term $[g_RV_RL_0]$, and obtain $L_0=0$ by the equation of motion of $V_R$. This leads to the original action~(41). Therefore, the old minimal SUGRA action~(40) cannot be embedded into the new-minimal SUGRA formulation.

We would like to thank the anonymous referee for pointing out this critical error in our statement.

\appendix
\section{proof}
Let us consider a general old-minimal SUGRA action with U(1)$_R$ but without a kinetic term of the U(1)$_R$ vector superfield,
\begin{align}
S=\left[\frac{1}{2}\tilde{S}_0\overline{\tilde{S}}_0e^{-g_RV_R}\Omega(\tilde{S}^i,\overline{\tilde{S}}^{\bar{j}}e^{-g_Rm_jV_R},\Phi,\overline{\Phi})\right]_D+[\tilde{S}_0^3\tilde{W}(\tilde{S}^i,\Phi)]_F\label{gen},
\end{align}
where $\tilde{S}_0$, $\tilde{S}^i$, and $\Phi$ denote a compensator, matter, Stuckelberg  chiral superfield, respectively, and they are transformed under U(1)$_R$ as,
\begin{align}
&\tilde{S}_0\to\tilde{S}_0e^{i\Lambda},\\
&\tilde{S}^i\to\tilde{S}^ie^{im_i\Lambda},\\
&\Phi\to\Phi+i\Lambda,
\end{align}
where $\Lambda$ is a chiral superfield. By the following field redefinitions, we make the compensator and matter as singlet under U(1)$_R$,
\begin{align}
S_0\equiv \tilde{S}_0e^{-\Phi},\quad S^i\equiv \tilde{S}^ie^{-m_i\Phi}.\nonumber
\end{align}
Then, the action~(\ref{gen}) becomes
\begin{align}
S=\left[\frac{1}{2}S_0\overline{S}_0\Omega(S^i,\overline{S}^{\bar{j}}e^{-g_Rm_jV_R},\Phi+\overline{\Phi}+g_RV_R)\right]_D+[S_0^3W(S^i)]_F.
\end{align}
Note that, due to the U(1)$_R$ gauge invariance, U(1)$_R$ vector superfield only appear in the combination $\Phi+\overline{\Phi}+g_RV_R$. Using a Lagrange multiplier $U$, we can rewrite this action as
\begin{align}
S=\left[\frac{1}{2}S_0\overline{S}_0\Omega(S^i,\overline{S}^{\bar{j}},U)\right]_D+[S_0^3W(S^i)]_F+[L_0(U-g_RV_R)]_D,
\end{align}
where $U$ is a real general superfield, and $L_0$ is a real linear superfield. The variation of $L_0$ gives
\begin{align}
U=\Phi+\overline{\Phi}+g_RV_R,
\end{align}
which reproduces the previous action. Instead, the variation of $U$ gives 
\begin{align}
L_0+\frac{1}{2}S_0\overline{S}_0\partial_U\Omega=0,
\end{align} 
which, in principle, is solved with respect to $U$ and yields $U=U(S_0, \overline{S}_0, S^i,\overline{S}^{\bar{j}}, L_0)$. Note that $U$ does not depend on $V_R$. Then, substituting $U=U(S_0, \overline{S}_0, S^i,\overline{S}^{\bar{j}}, L_0)$, we obtain
\begin{align}
S=[{\cal F}(S_0, \overline{S}_0, S^i, \overline{S}^{\bar{j}}, L_0)]_D+[S_0^3W(S^i)]_F+[-g_RV_RL_0]_D,
\end{align}
where
\begin{align}
{\cal F}(S_0, \overline{S}_0, S^i, \overline{S}^{\bar{j}}, L_0)\equiv \left(\frac{1}{2}S_0\overline{S}_0\Omega(S^i,\overline{S}^{\bar{j}},U)+L_0U\right)|_{U=U(S_0, \overline{S}_0, S^i,\overline{S}^{\bar{j}}, L_0)}.
\end{align}
The action has $V_R$ only in the last term, which leads to $L_0=0$. Thus, we find that the gauged R old-minimal action~(\ref{gen}) cannot be the new-minimal SUGRA action. Note that this conclusion comes from the assumption that $V_R$ does not have a kinetic term, and if there exists such a term, we do not obtain $L_0=0$.

\end{document}